\documentclass{article} 
\PassOptionsToPackage{numbers, compress}{natbib}
\usepackage[preprint]{neurips_2021}
\usepackage{hyperref}
\usepackage{url}
\usepackage{amsmath,amssymb,amsfonts}
\usepackage{bm}
\usepackage{algorithmic}
\usepackage{algorithm}
\renewcommand{\algorithmiccomment}[1]{\bgroup\hfill//~#1\egroup}
\usepackage{graphicx}
\usepackage{subcaption}
\newcommand*\system{\textsc{FedSIA}}
\usepackage{tabularx}
\usepackage{multirow}
\usepackage{booktabs}
\newtheorem{definition}{Definition}
\newtheorem{theorem}{Theorem}

\title{Source Inference Attacks in Federated Learning}

\author{%
  Hongsheng Hu \\
  University of Auckland \\
  New Zealand \\
  \texttt{hhu603@aucklanduni.ac.nz} \\
   \And
   Zoran Salcic \\
   University of Auckland \\
   New Zealand \\
   \texttt{z.salcic@auckland.ac.nz} \\
   \And
   Lichao Sun \\
   Lehigh University \\
   USA \\
   \texttt{lis221@lehigh.edu} \\
   \AND
   Gillian Dobbie \\
   University of Auckland \\
   New Zealand \\
   \texttt{g.dobbie@auckland.ac.nz} \\
   \And
   Xuyun Zhang\thanks{Corresponding author.} \\
   Macquarie University \\
   Australia \\
  \texttt{xuyun.zhang@mq.edu.au} \\
  
}

\begin{document}

\maketitle

\begin{abstract}
Federated learning (FL) has emerged as a promising privacy-aware paradigm that allows multiple clients to jointly train a model without sharing their private data. Recently, many studies have shown that FL is vulnerable to \textit{membership inference attacks} (MIAs) that can distinguish the training members of the given model from the non-members. However, existing MIAs ignore the source of a training member, \textit{i.e.}, the information of which client owns the training member, while it is essential to explore source privacy in FL beyond membership privacy of examples from all clients. The leakage of source information can lead to severe privacy issues. For example, identification of the hospital contributing to the training of an FL model for COVID-19 pandemic can render the owner of a data record from this hospital more prone to discrimination if the hospital is in a high risk region. In this paper, we propose a new inference attack called \textit{source inference attack} (SIA), which can derive an optimal estimation of the source of a training member. Specifically, we innovatively adopt the Bayesian perspective to demonstrate that an honest-but-curious server can launch an SIA to steal non-trivial source information of the training members without violating the FL protocol. The server leverages the prediction loss of local models on the training members to achieve the attack effectively and non-intrusively. We conduct extensive experiments on one synthetic and five real datasets to evaluate the key factors in an SIA, and the results show the efficacy of the proposed source inference attack.
\end{abstract}

\section{Introduction}
Big data and deep learning technologies have enabled us to perform scalable data mining across multiple parties to build powerful prediction models. For example, it will be very appealing for different countries to collaborate, utilizing their medical data records to train prediction models for fighting against the COVID-19 pandemic. However, many countries or regions have issued strong privacy protection laws and regulations, such as GDPR \cite{regulation2016regulation}, and it is very difficult to straightforwardly collect and combine the data from different parties for a data mining task. To circumvent this major obstacle towards big data mining, a novel machine learning (ML) paradigm named feaderated learning  (FL) has  recently been proposed, which allows multiple clients coordinated by a central server to train a joint ML model in an iterative manner~\cite{mcmahan2017communication,he2020fedml,he2021fedgraphnn}. In FL, no client can access any training data owned by other clients, leading to a privacy-aware paradigm for collaborative model training. Specific to the example mentioned above, FL can greatly facilitate the scenario where many hospitals hope to build a joint COVID-19 diagnosis ML model from their distributed data. A real-life case has been shown in~\cite{xu2020collaborative}, where FL has been successfully adopted to build a promising ML model for COVID-19 diagnosis, with the use of the geographically distributed chest CT (Computed Tomography) data collected from patients at different hospitals.

However, many recent studies~\cite{melis2019exploiting,nasr2019comprehensive,zhudeep,wang2019beyond,truex2018towards} have demonstrated that FL fails to provide sufficient privacy
guarantees, as sensitive information can be revealed in the training process. In FL, multiple clients send ML model weight or gradient updates derived from local training to a central server for global model training. The communication of model updates renders FL vulnerable to several recently developed privacy attacks, such as property inference attacks~\cite{ganju2018property}, reconstruction attacks~\cite{hitaj2017deep}, and membership inference attacks (MIAs)~\cite{shokri2017membership}. Among these attacks, MIAs  aim to identify whether or not a data record was in the training dataset the model was built on ({\em i.e.,} a member). This can impose severe privacy risks on individuals. For example, via identifying the fact that a clinical record that has been used to train a model associated with a certain disease,  MIAs can infer that the owner of the clinical record has a high chance of having the disease. 

However, existing MIAs ignore the source of a training member, i.e., the information of which client owns the training member, while MIAs against FL models distinguish the training members of the model from the non-members. It is essential to explore source privacy in FL beyond membership privacy, because the leakage of such information can lead to further privacy issues. For instance, in the scenario where multiple hospitals jointly train an FL model for COVID-19 diagnosis, MIAs can only reveal who have been tested for COVID-19, but the further identification of the source hospital where the people are from will make them more prone to discrimination, especially when the hospital is in a high risk region or country~\cite{devakumar2020racism}.

In this paper, we propose a novel inference attack called \textit{Source Inference Attack} (SIA) in the context of FL. SIA aims to determine which client owns a training record in FL. In practice, the SIA can be considered as a natural extension beyond MIAs, i.e., after determining which data instances are training members in MIAs, the adversary can further conduct the SIA to identify which client it comes from. To be practical, it is assumed that the adversary is an \textit{honest-but-curious} central server, who knows the identities of clients and receives the updates from them. It is worth noting that the server can infer client-private information without interfering with the FL training nor affecting the model prediction performance. While the adversary can be one of the clients, we argue that it is impractical for her to launch SIAs as she knows little about other clients' identities and can only access the joint models.

Specifically, we innovatively explore the SIA from the Bayesian perspective, and demonstrate that a server can achieve the optimal estimate of the source of a training member in an SIA without violating the FL protocol. To this end, the prediction loss of local models on the training members is utilised to obtain the source information of the training members effectively and non-intrusively. Besides theoretical formulation, we empirically evaluate the SIA in FL trained with one synthetic and five real world datasets, with respect to several FL aspects such as data distributions across clients, the number of clients, and the number of local epochs. The experiment results validate the efficacy of the proposed source inference attack under various FL settings. An important finding is that the success of an SIA is directly relevant to the generalizability of local models and the diversity of the local data. 

Our main contribution is multifold, summarized as follows. 
\begin{itemize}
    \item First, we propose the source inference attack (SIA), a novel inference attack in FL that identifies the source of a training member. SIA can further breach the privacy of training members beyond membership inference attacks. 
    \item Second, we adopt the Bayesian perspective to demonstrate that an honest-but-curious central server can fulfil an effective SIA in a non-intrusive manner by optimally estimating the source of a training member, using prediction loss of local models. 
    \item Last, we perform an extensive empirical evaluation on both synthetic and real world datasets under various FL settings, and the results validate the efficacy of the proposed SIA. 
\end{itemize}
We provide all proofs in the full version of our paper, and our source code is available at: \url{https://github.com/HongshengHu/source-inference-FL}.

\section{Preliminaries}~\label{2}
In this section, we briefly review the background of the federated learning and  membership inference attacks.

\subsection{Federated Learning}
Federated learning allows multiple clients to jointly train an ML  model in an interactive manner. It is an attractive framework for training ML models without direct access to diverse training data owned by different clients, especially for privacy-sensitive tasks~\cite{mcmahan2017communication,zhao2018federated,melis2019exploiting,bagdasaryan2020backdoor}. The \textit{federated averaging} (FedAvg)~\cite{mcmahan2017communication} algorithm is the first and perhaps the most widely used FL algorithm. During multiple rounds of communication between server and clients, a central model is trained. At each communication round, the server distributes the current central model to local clients. The local clients then perform local optimization using their own data. To minimize communication, clients might update the local model for several epochs during a single communication round. Next, the optimized local models are sent back to the server, who average them to allocate a new central model. The performance of the new central model decides the training is either stopped or a new communication round starts. In FL, clients never share data, only their model weights or gradients. 

\subsection{Membership Inference Attacks}
Membership inference attacks aim to identify whether a data record was part of the target model's training dataset or not. Shokri et al.~\cite{shokri2017membership} present the first MIAs against ML models. Specifically, they demonstrate that an adversary can tell whether a data record has been used to train a classifier or not, solely based on the prediction vector of the data record. Since then, a growing body of work further investigates and explores the feasibility of MIAs on various ML models~\cite{hu2021membership}. Nevertheless, recent works~\cite{melis2019exploiting, nasr2019comprehensive} have demonstrated the success of MIAs on FL models. For example, Melis et al.~\cite{melis2019exploiting} have shown that an adversary can infer whether a specific location profile was used to train an FL model on the FourSquare location dataset with high success rate. Although MIAs can distinguish the training members of the FL model from the non-members, existing inference attacks ignore to further explore which client owns the training member identified by MIAs. In this paper, we fill this gap and show the possibilities of breaching the source privacy of training members. 

\section{Source Inference Attacks}\label{3}
In this section, we formally analyze how an honest-but-curious sever in FL can optimally estimate the source of a training member from the Bayesian perspective. 

We focus on the supervised learning of classification tasks. The adversary is the honest-but-curious server who faithfully implements FedAvg while trying to determine where a training data record comes from. Assuming the whole training dataset consists of $n$ i.i.d. data records ${\bm{z}_1, \cdots, \bm{z}_n}$ from a data distribution. Each record is represented as $\bm{z}=(\bm{x},y)$ where $\bm{x}$ is an input vector and $y$ is the class label. The source status of each record is represented by a $K$-dimensional (assuming there are $K$ clients) multinomial vector $\bm{s}$ in which one of the elements $s_k$ equals $1$, and all remaining elements equal $0$. We assume that multinomial source variables $\bm{s}_1, \cdots, \bm{s}_n$ are independent, and the training record $\bm{z}_i$ comes from the client k with the probability $\mathbb{P}({s}_{ik}=1)=\lambda$. Without loss of generality, taking the case of $\bm{z}_1$, the source inference is defined as follows:

\begin{definition}[Source inference]
\label{def1}
Given local optimized model $\bm{\theta}_k$, a training record $\bm{z}_1$, source inference aims to infer the posterior probability of $\bm{z}_1$ belonging to the client k:
\begin{equation}
    \mathcal{S}({\bm{\theta}_k},{\bm{z}_1}): = \mathbb{P}({{s}_{1k}} = 1|{\bm{\theta}_k},{\bm{z}_1}).
\end{equation}
\end{definition}

For the source inference by \textit{Definition~\ref{def1}}, we want to derive the explicit formula for $\mathcal{S}({\bm{\theta}_k},{\bm{z}_1})$ from the Bayesian perspective, which establishes the optimal limit that our source inference can achieve. We denote $\bm{\tau}=\{\bm{z}_2,\cdots,\bm{z}_n,\bm{s}_2,\cdots,\bm{s}_n\}$ as the set which collects the knowledge about the other training records and their source status. The explicit formula of $\mathcal{S}({\bm{\theta}_k},{\bm{z}_1})$ is given by the following theorem.

\begin{theorem}
Given local optimized model $\bm{\theta}_k$, a training record $\bm{z}_1$, the optimal source inference is given by:
\begin{equation}
    \mathcal{S}({\bm{\theta}_k},{\bm{z}_1}) = {\mathbb{E}_{\bm{\tau}} }\left[ {{\sigma} \left( {\log (\frac{{\mathbb{P}({\bm{\theta} _k}|{{s}_{1k}} = 1,{\bm{z}_1},\bm{\tau} )}}{{\mathbb{P}({\bm{\theta} _k}|{{s}_{1k}} = 0,{\bm{z}_1},\bm{\tau} )}}) + {\mu _\lambda }} \right)} \right],
\end{equation}
\end{theorem}
where $\mu_{\lambda}=\log (\frac{\lambda }{{1 - \lambda }})$ and ${\sigma}(\cdot)$ is the sigmoid function.


We observe that \textit{Theorem 1} does not have the loss $\ell(\cdot)$ form and only relies on the posterior parameter $\bm{\theta}_k$ in expectation given $\{\bm{z}_1,\cdots,\bm{z}_n,\bm{s}_1,\cdots,\bm{s}_n \}$ is a random variable. To make $\mathcal{S}({\bm{\theta}_k},{\bm{z}_1})$ more explicit with the loss term, we assume an ML algorithm produced parameters $\bm{\theta}$ follows a posterior distribution. According to energy based models~\cite{lecun2006tutorial,du2019}, the posterior distribution of an ML model $\bm{\theta}$ follows:
\begin{equation}
    p(\bm{\theta} |{\bm{z}_1}, \cdots ,{\bm{z}_n}) \propto {e^{ - \frac{1}{\gamma }\sum\nolimits_{i = 1}^n {\ell (\bm{\theta} ,{\bm{z}_i})} }},
\end{equation}
where $\gamma$ is a temperature parameter controlling the stochasticity of $\bm{\theta}$. Following this assumption, given $\{\bm{z}_1,\cdots,\bm{z}_n, \bm{s}_1, \cdots,\bm{s}_n \}$, the posterior distribution of $\bm{\theta}_k$ follows:
\begin{equation}
    p({\bm{\theta} _k}|{\bm{z}_1}, \cdots ,{\bm{z}_n},{\bm{s}_1}, \cdots ,{\bm{s}_n}) \propto {e^{ - \frac{1}{\gamma }\sum\nolimits_{i = 1}^n {{s_{ik}}\ell (\bm{\theta}_k ,{\bm{z}_i})} }}.
\end{equation}
We further define the posterior distribution of $\bm{\theta}_k$ given training samples $\bm{z}_2,\cdots,\bm{z}_n$ and their source status $\bm{s}_2,\cdots,\bm{s}_n$ (i.e., given $\bm{\tau}$):
\begin{equation}
    {p_{\bm{\tau}} }({\bm{\theta} _k}): = \frac{{{e^{ - \frac{1}{\gamma }\sum\nolimits_{i = 2}^n {{s_{ik}}\ell (\bm{\theta}_k ,{\bm{z}_i})} }}}}{{\int_{\bm{t}} {{e^{ - \frac{1}{\gamma }\sum\nolimits_{i = 2}^n {{s_{ik}}\ell (\bm{t} ,{\bm{z}_i})} }}} d\bm{t}}},
\end{equation}
where the denominator is a constant. The following theorem explicitly demonstrates how to conduct the optimal source inference with the loss term.
\begin{theorem}
Given a local resulting model $\bm{\theta}_k$, a training record $\bm{z}_1$, the optimal SIA is given by:
\begin{equation}
    \mathcal{S}({\bm{\theta} _k},{\bm{z}_1}) = {\mathbb{E}_{\bm{\tau}} }\left[ {\sigma \left( {g({\bm{z}_1},\bm{\theta} ,{p_{\bm{\tau}} }) + {\mu _\lambda }} \right)} \right],
\end{equation}
where
\begin{align}
{\ell _{{p_{\bm{\tau}} }}}({\bm{z}_1}): &=  - \gamma \log \left( {\int_{\bm{t}} {{e^{ - \frac{1}{\gamma }\ell (\bm{t},{\bm{z}_1})}}} {p_{\bm{\tau}} }(\bm{t})d\bm{t}} \right) \label{loss_other},\\
\ell (\bm{\theta}_k,{\bm{z}_1}): &=  - \gamma \log \left( {{e^{ - \frac{1}{\gamma }\ell ({\bm{\theta}_k},{\bm{z}_1})}}} \right) \label{loss_k}, \\
g({\bm{z}_1},\bm{\theta} ,{p_{\bm{\tau}} }): &= \frac{1}{\gamma }({\ell _{{p_{\bm{\tau}} }}}({\bm{z}_1}) - \ell ({\bm{\theta} _k},{\bm{z}_1})). \label{gap}
\end{align}
\end{theorem}


The term $g({\bm{z}_1},\bm{\theta} ,{p_{\bm{\tau}} })$ in Equation~\ref{gap} is the gap between $\ell _{{p_{\bm{\tau}} }}({\bm{z}_1})$ and $\ell \left({\bm{\theta} _k},{\bm{z}_1})\right)$. Since $\bm{\tau}$ is a training set that does not contain any information about $\bm{z}_1$, $p_{\bm{\tau}}$ corresponds to a posterior distribution of the parameters of an ML model that was trained without seeing $\bm{z}_1$. Note that $\ell \left({\bm{\theta} _k},{\bm{z}_1}\right)$ is the local model $\bm{\theta}_k$'s evaluation of the loss on the training record $\bm{z}_1$. Comparing Equation~\ref{loss_other} and Equation~\ref{loss_k}, we can easily find that $\ell _{{p_{\bm{\tau}} }}({\bm{z}_1})$ is the expectation of the loss $\ell (\cdot,\bm{z}_1)$ over the typical models that have not seen $\bm{z}_1$. Thus, we can interpret $g({\bm{z}_1},\bm{\theta} ,{p_{\bm{\tau}} })$ as the difference between $\bm{\theta}_k$'s loss on $\bm{z}_1$ and other models' (trained without $\bm{z}_1$) average loss on $\bm{z}_1$.

In FL, the malicious server can implement an SIA in each communication round. The server receives the updated local models from each client and conducts the SIA to identify whether $\bm{z}_1$ belongs to the client k. Let us qualitatively analyze $\mathcal{S}({\bm{\theta} _k},{\bm{z}_1})$ in \textit{Theorem 2}. $\mathcal{S}({\bm{\theta} _k},{\bm{z}_1})$ has two important terms $g({\bm{z}_1},\bm{\theta} ,{p_{\bm{\tau}} })$ and ${\mu _\lambda }$, which decide the posterior probability. In FL, $\ell \left({\bm{\theta} _k},{\bm{z}_1}\right)$ represents the local updated model $\bm{\theta}_k$'s loss on $\bm{z}_1$. $\ell _{{p_{\bm{\tau}} }}({\bm{z}_1})$ represents the average loss of $\bm{z}_1$ under the local models $\bm{\theta}_1,\cdots,\bm{\theta}_{k-1},\bm{\theta}_{k+1},\cdots,\bm{\theta}_K$ that are updated without $\bm{z}_1$. Note that $\ell(\cdot)$ is a loss function which measures the performance of a model on a data record. If $\ell _{{p_{\bm{\tau}} }}({\bm{z}_1}) \approx \ell \left({\bm{\theta} _k},{\bm{z}_1}\right)$, which means the client k behaves almost the same as other clients on $\bm{z}_1$, then $g({\bm{z}_1},\bm{\theta} ,{p_{\bm{\tau}} }) \approx 0$. Since $\sigma(\mu_{\lambda})=\lambda$, the posterior probability $\mathcal{S}({\bm{\theta} _k},{\bm{z}_1})$ is equal to $\lambda$. Thus, we have no information gain on $\bm{z}_1$ beyond prior knowledge. In FL, the prior knowledge is $\mathbb{P}({s}_{ik}=1)=\lambda=\frac{1}{K}$. In this case, the source inference is equal to a \textit{random guess}. However, if $\ell _{{p_{\bm{\tau}} }}({\bm{z}_1}) > \ell \left({\bm{\theta} _k},{\bm{z}_1}\right)$, that is, the client k performs better than other clients on $\bm{z}_1$, $g({\bm{z}_1},\bm{\theta} ,{p_{\bm{\tau}} })$ becomes positive. When $g({\bm{z}_1},\bm{\theta} ,{p_{\bm{\tau}} })>0$, $\mathbb{P}({{s}_{1k}} = 1|{\bm{\theta}_k},{\bm{z}_1})>\lambda$ and thus we gain non-trivial source information on $\bm{z}_1$. Moreover, since $\sigma(\cdot)$ is non decreasing, smaller $ \ell \left({\bm{\theta} _k},{\bm{z}_1})\right)$ indicates a higher probability that $\bm{z}_1$ belonging to the client k. 

\begin{algorithm}[t!]
\caption{\system\: The $K$ clients are indexed by $k$; $B$ is the local mini-batch size; $E$ is the number of local epochs; $\eta$ is the learning rate; $\bm{z}_1$ is a training data.}
\label{fedsia}
\begin{algorithmic}[1]
\STATE \textbf{Server executes}

\STATE initialize $\theta_{0}$ \COMMENT{initialize weights}
\STATE $m \leftarrow max (C \cdot K,1)$
\FOR{each round $t = 1$ to $T$}
\STATE $S_t \leftarrow $ (random set of m clients)
\FOR{each client $k \in S_t$}
\STATE $\theta^{k}_{t}$ $\leftarrow$ \textbf{ClientUpdate}($\theta_{t-1}^{k}$)
\STATE Compute $\ell_{k}({\theta^{k}_{t},\bm{z}_1)}$ \COMMENT{calculate local loss on $\bm{z}_1$}
\ENDFOR 

\STATE $i \leftarrow argmin (\ell_{1}(\bm{\theta}_1,\bm{z}_1) ,\cdots,\ell_{m}(\bm{\theta}_m,\bm{z}_1))$ \COMMENT{source}

\STATE ${\theta _t} \leftarrow \sum\nolimits_k {\frac{{{n^{(k)}}}}{n}} \theta _t^k$ \COMMENT{update central model}
\ENDFOR

\STATE \textbf{ClientUpdate}($\theta$)
\begin{ALC@g}
\STATE  $\mathcal{B}$ $\leftarrow$ (split $D_{k}$ into batches of size $B$) 
\FOR{each local epoch $i$ from $1$ to $E$}
\FOR{batch $b \in \mathcal{B}$}
\STATE $\theta  \leftarrow \theta  - \eta \nabla \ell (b;\theta )$ \COMMENT{mini-batch gradient descent}
\ENDFOR
\ENDFOR
\RETURN $\theta$ \COMMENT{return model to central server}
\end{ALC@g}
\end{algorithmic}
\end{algorithm}

We conclude that the smaller loss of client k's local model on a training record $\bm{z}_1$, the higher posterior probability that $\bm{z}_1$ belongs to the client k. This motivates us to design the SIA in FL such that the client whose local model has the smallest loss on a training record should own this record. Moreover, if the client's local model's behavior on its local training data is different from that of other clients, our attack will always achieve better performance than random guess. We give more empirical evidence in Section~\ref{experiments}. Based on the conclusion above, we propose \system\ as described in Algorithm 1, an FL framework based on FedAvg~\cite{mcmahan2017communication} that allows an honest-but-curious server to implement SIAs without violating the FedAvg protocol. 

\section{Experiments}\label{experiments}\label{4}
\subsection{Datasets and Model Architectures}
In the experiments, we evaluate SIAs on six datasets, i.e., Synthetic, Location\footnote{https://sites.google.com/site/yangdingqi/home/foursquare-dataset}, Purchase\footnote{https://www.kaggle.com/c/acquire-valued-shoppers-challenge/data}, CHMNIST\footnote{https://www.kaggle.com/kmader/colorectal-histology-mnist}, MNIST\footnote{http://yann.lecun.com/exdb/mnist/}, and CIFAR-10\footnote{https://www.cs.toronto.edu/~kriz/cifar.html}. Among them, Synthetic is a synthetic i.i.d. dataset, which allows us to manipulate data heterogeneity more precisely. We follow the same generation setup as described in~\cite{LiSZSTS20,li2019convergence}. Location, Purchase, CHMNIST, MNIST, and CIFAR-10 are realistic datasets which are widely used for evaluating privacy leakage on ML models~\cite{shokri2017membership,jayaraman2019evaluating,ganju2018property,wang2019beyond}. For MNIST and CIFAR-10, we use the training dataset and testing dataset given. For the rest of the datasets, we randomly select $80$\% samples as the training records and use the remaining $20$\% samples as the testing records.

We consider deep neural networks (DNN) as the collaborative models for the classification tasks. In particular, for MNIST, CHMNIST, CIFAR-10, we use a convolutional neural network with two 5x5 convolution layers (the first with 32 channels, the second with 64, each followed with 2x2 max pooling), two fully connected layers with 512 and 128 units and ReLu activation, and a final softmax output layer. For Synthetic, Location, and Purchase, we use a fully-connected neural network with 1-hidden layer with 200 units each using ReLu activations. For each client in FL, we set a local mini-batch size of $12$ for all the experiments. For all models, we use SGD with the learning rate of $0.01$. Our DNN architecture does not necessarily achieve the highest classification accuracy for the considered datasets, as our goal is not to attack the best DNN architecture. Our goal is to show that SIAs can identify which local client a training record comes from when the DNN classifier is trained in a federated manner. 

In our experiment, we randomly select $100$ training records from each client as the target training examples of which the server wants to identify the source. We set the fraction of the clients $C$ to 1 in FL to simplify our experiments as we ignore the efficiency of the FL training when analyzing the privacy leakage. We consider \textit{attack success rate} (ASR) as the evaluation metric for the source inference. The ASR is defined as the fraction of the target records' where the source status are correctly identified by the server. We consider a trivial attack of \textit{randomly guessing}, which randomly selects a client as the source of the target training record as the performance baseline of an SIA. For all the learning tasks, we train the central model for $20$ rounds, which is enough for the central model to converge. We record ASR during each communication round and report the highest ASR. All experiments are implemented using PyTorch with a single GPU NVIDIA Tesla P40. 

\subsection{Factors in Source Inference Attack}
\noindent\textbf{Data Distribution.} The training data across clients are usually non-i.i.d. (heterogeneity) in FL. That is, a client's local data can not be regarded as samples drawn from the overall data distribution. If the training data is more heterogeneous, each local optimized model will be more different during the FL training, which benefits SIAs. Intuitively, an SIA is more effective when the degree of data heterogeneity increases. To simulate heterogeneity of training data, we follow the method used in~\cite{xie2019dba,bagdasaryan2020backdoor,lin2020ensemble} and use a Dirichlet distribution to divide the training records. The degree of data heterogeneity is controlled by a hyperparameter $\alpha$ ($\alpha>0$) of the Dirichlet distribution. In general, the reverse of the magnitude of $\alpha$ reflects the degree of data heterogeneity. 

\noindent\textbf{Number of Local Epochs.} In each communication round of FL, the client locally runs SGD on the current central model using its entire training dataset for several epochs and then submits the optimized model to the server. Recent studies~\cite{song2017machine,carlini2019secret} have demonstrated that ML models are prone to memorize their training data. Intuitively, if a client updates the model on its local dataset with more epochs in each communication round, its local resulting model remembers the information of the local dataset better, which benefits SIAs. 

\begin{figure}
    \centering
    \includegraphics[width=0.6\linewidth]{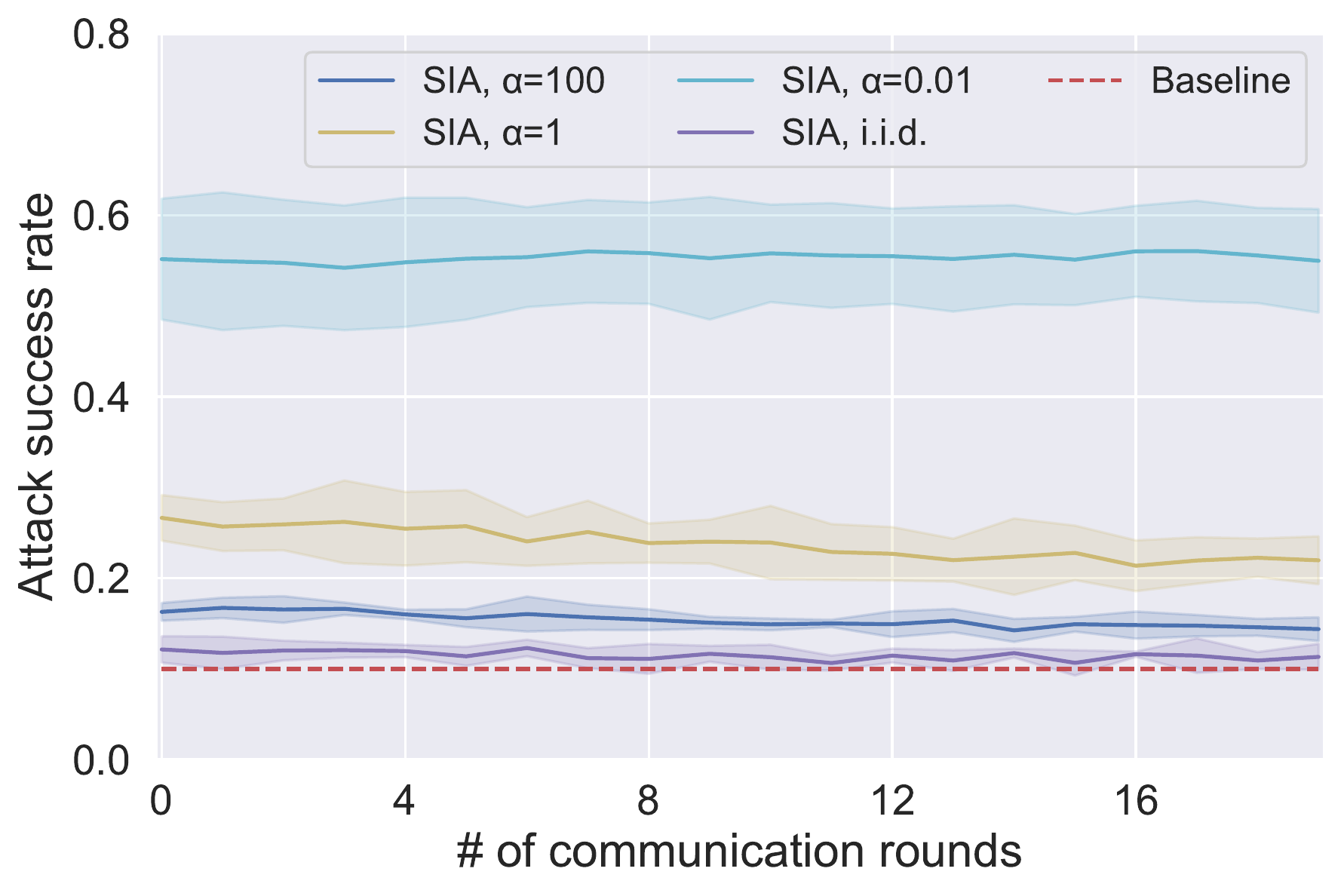}
    \caption{ASR of source inference on the Synthetic dataset in various data distribution settings. The x-axis represents the number of communication rounds. The y-axis  represents ASR.}
  \label{att_synthetic} 
\end{figure}

\subsection{Source Inference on Synthetic Dataset}
We first conduct experiments on Synthetic to investigate how data distribution affects SIAs, because synthetic data allows us to manipulate the heterogeneity of the training data precisely. Without loss of generality, we assume there are $10$ clients and $E=5$. Fig.~\ref{att_synthetic} depicts the ASR of SIAs in each communication round during the FL training. We observe that our proposed SIAs always perform better than the random guessing baseline. This serves as empirical evidence for our theoretical analysis that random guess is the lower bound of our optimal source inference. The attacker performs better when the local data changes from i.i.d. to non-i.i.d., and the ASR increases as the heterogeneity of data increases.

\subsection{Source Inference on All Dataset}\label{4D}
We have demonstrated that SIAs are effective on synthetic data in both i.i.d. and non-i.i.d. settings. Now we use real datasets to further validate the effectiveness of SIAs and investigate the factors affecting the performance of SIAs. The SIAs leverages the local models' different prediction loss on the training examples. Intuitively, if the local model is overfitted, it will perform much better on its training members than other data, i.e., distinguishable prediction loss between the local training data and other data. We link the level of non-i.i.d. and the number of local epochs to overfitting to study how the two factors affect the performance of SIAs.

\begin{figure}
    \centering
    \subfloat[SIAs under various $\alpha$.]{\includegraphics[width=0.45\textwidth]{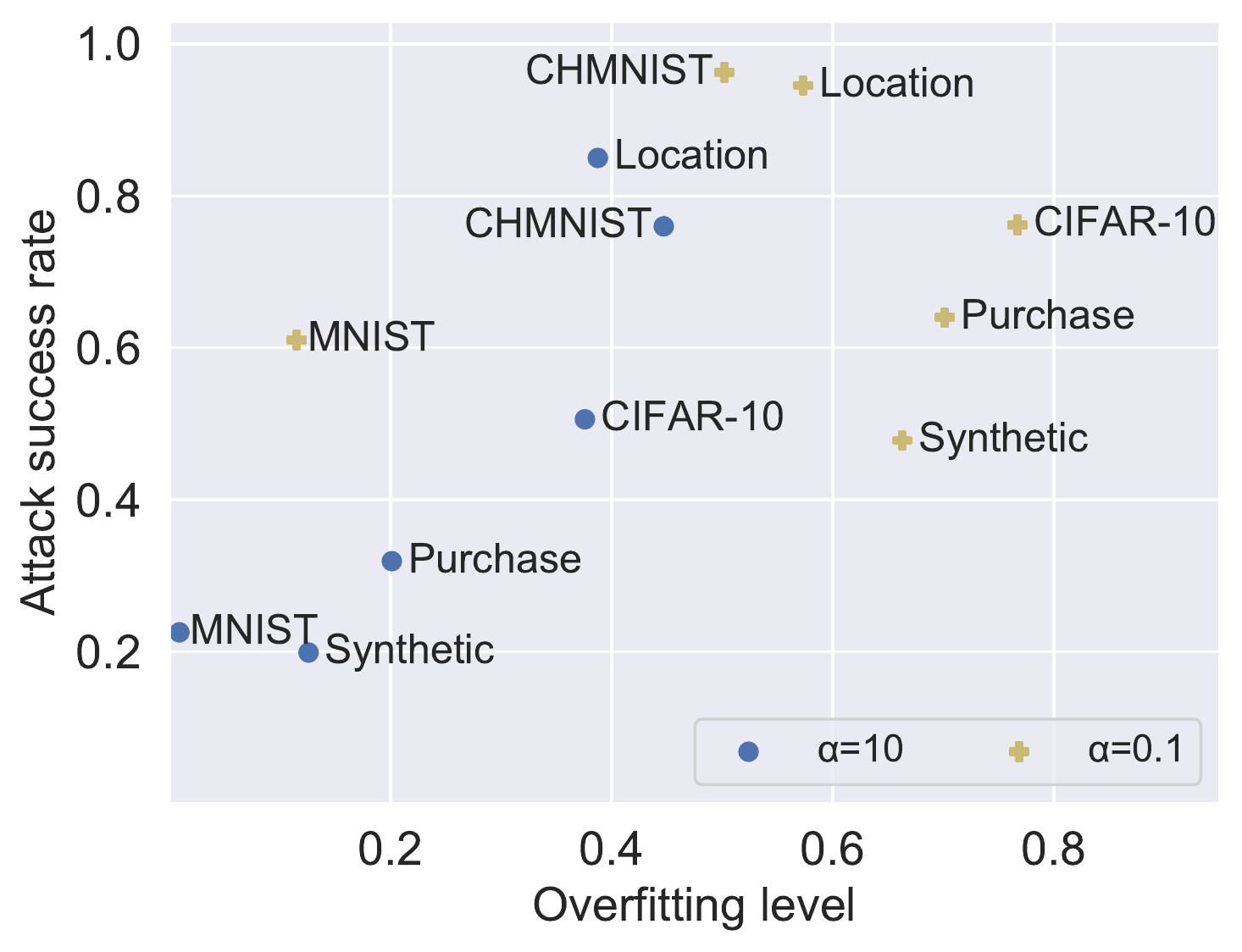}}
    \hspace{10pt}
    \subfloat[SIAs under various $E$.]{\includegraphics[width=0.45\textwidth]{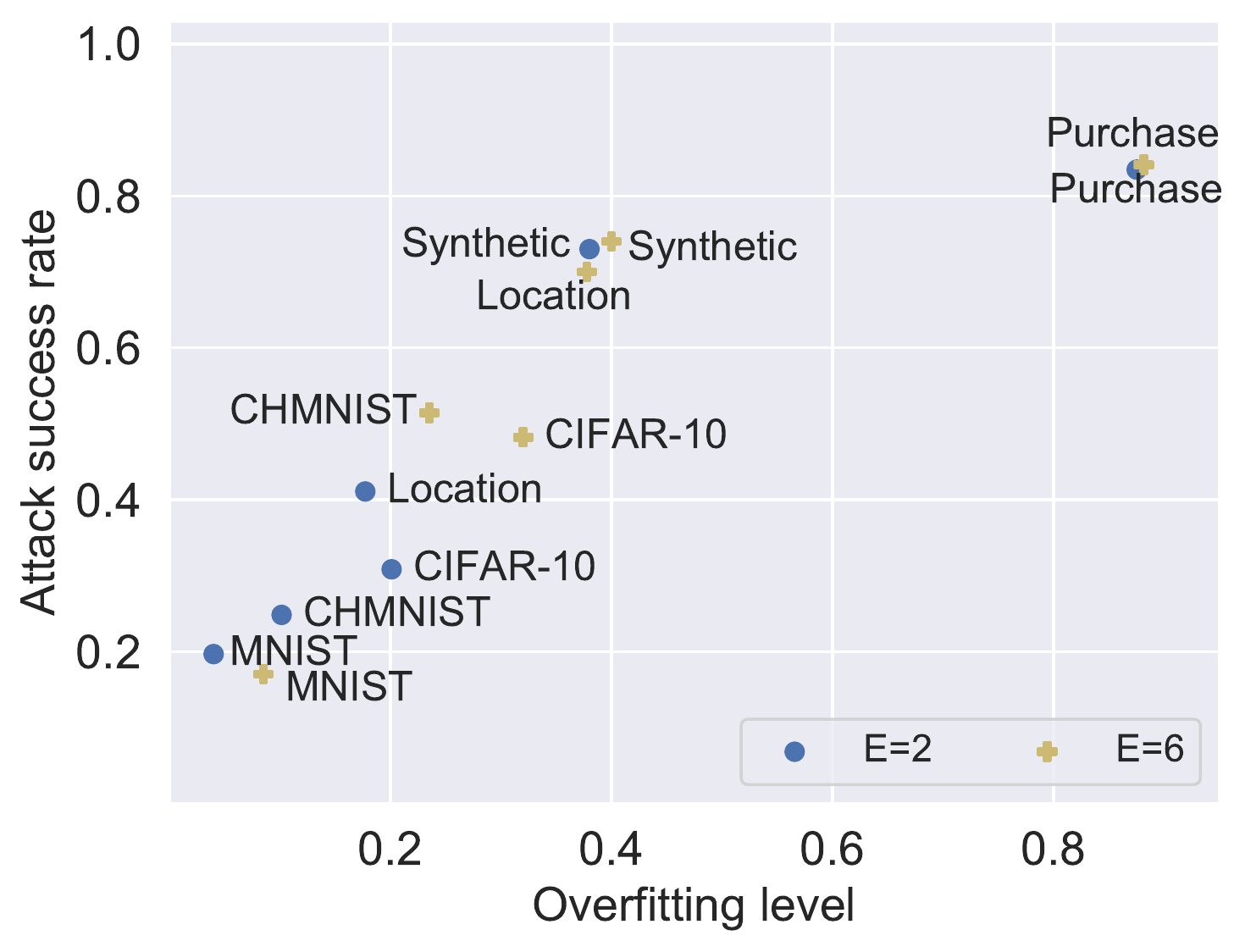}}
    \caption{The overfitting levels affect the performance of SIAs in FL, where the x-axis represents different overfitting levels and the y-axis represents ASR. (a) We fix $E$ and $K$ for the FL model on the same dataset and only change $\alpha$ from $10$ to $0.1$. (b) We fix $K$ and $\alpha$ for the FL model on the same dataset and only change $E$ from $2$ to $6$.}
    \label{overfit}
\end{figure}

Fig.~\ref{overfit} shows the SIAs' ASR of different FL models under different overfitting levels. The overfitting level of the FL model here is calculated as the average of all local models' generalization gap. As we can see, increasing the level of non-i.i.d. across clients (i.e., changing $\alpha$ from $10$ to $0.1$) increases the ASR of SIAs in all models, as increasing the level of non-i.i.d. will inevitably increase the level of overfitting. However, when we increase the number of local epochs from $2$ to $6$, the ASRs of SIAs on CHMNIST, CIAFR-10, Location increases while on MNIST, Synthtic, Purchase the ASRs does not vary much. This is because changing local epochs does not increase the overfitting level of all models.

Success of SIAs is directly related to the generalizability of the local models and the diversity of the local training data. If a local model generalizes well to inputs beyond its local training members, it will not leak too much source information about its local data. Moreover, if the local training set fails to represent the overall training data distribution, the local model leaks significant information about its local data and the ASR of SIAs remains high. Recent works~\cite{zhao2018federated,li2019convergence,li2020federated} have demonstrated that the non-i.i.d. of training data in FL has brought statistical heterogeneity challenges for model convergence guarantees. In this paper, we show another harm of non-i.i.d.: the leakage of source privacy for local data.

\vspace{-10pt}
\section{Discussion}~\label{5}
\vspace{-10pt}

\begin{table}[t!]
\caption{Source Inference Defense via Differential Privacy.}
\label{dp}
\begin{center}
\resizebox{\textwidth}{!}{
\begin{tabular}{lcccccccc}
\toprule
\bfseries \multirow{2}{*}{Dataset} & \multicolumn{3}{c}{\bfseries FL without DP} & \multicolumn{4}{c}{\bfseries FL with DP} \\
\cline{2-8}
&$\text{Accuracy}_{\text{train}}$ & $\text{Accuracy}_{\text{test}}$ & $\text{ASR}_{\text{SIA}}$ & $\text{Accuracy}_{\text{train}}$ & $\text{Accuracy}_{\text{test}}$ & $\text{ASR}_{\text{SIA}}$ & $\epsilon$ \\
\hline
Location & 97.4\% & 71.2\% & 75.1\% & 12.7\% & 12.3\% &55.1\% & 1.31 \\
\hline
CIFAR-10 & 97.7\% & 68.3\% & 55.2\%  & 10.1\% & 10.0\% & 24.4\% & 1.48 \\
\bottomrule
\end{tabular}}
\end{center}
\vspace{-10pt}
\end{table}

Many works~\cite{shokri2015privacy,geyer2017differentially,mcmahan2017learning} suggest differential privacy~\cite{dwork2006calibrating} can be used against the inference attacks due to its theoretical guarantee of privacy protection. Here, we test the differential privacy as a defense technique against SIAs in FL. In this experiment, we evaluate the defense approaches on Location and CIFAR-10, as shown in Table~\ref{dp}.
In the experimental setting, we set $\alpha=10$, $K=2$, $E=5$ for Location, and $\alpha=10$, $K=5$, $E=5$ for CIAFR-10.
From the results, we can see that the ASRs drop from $75.1$\% to $55.1$\% on Location, and $55.2$\% to $24.4$\% on CIFAR-10, while applying differential privacy.
However, when differential privacy can defend the SIA,  it also hurts the performance of the model on its tasks, where the model utility drops from 71.2\% to 12.3\% on Location, and 68.3\% to 10.0\% on CIFAR-10. In this case, we can conclude that vanilla DP is not a effective solution for SIA in FL, which provides future research opportunities. 

\vspace{-10pt}
\section{Related work}\label{6}
\vspace{-5pt}


\subsection{Inference Attacks in FL}
Macahan et al. \cite{mcmahan2017communication} first propose the federated learning framework that can mitigate the privacy leakage of model training with limited, unbalanced, massively, or even non-IID data among distributed devices, such as mobile phones \cite{pan2021global}, healthcare data \cite{xu2021fedmood}.
The motivation is to share the model weights instead of the private data for better privacy protection.
However, recent works ~\cite{melis2019exploiting,nasr2019comprehensive,zhudeep,wang2019beyond,yang2020secure} investigate several privacy attacks in FL, including property inference attacks~\cite{ganju2018property}, reconstruction attacks~\cite{hitaj2017deep}, and membership inference attacks~\cite{shokri2017membership,wang2021membership}. MIAs in FL allows a malicious participant or server to distinguish the training members of the trained model from the non-members. Melis et al.~\cite{melis2019exploiting} first explore MIAs in FL and demonstrate that an adversary can infer whether a specific location profile was used to train an FL model on FourSquare location dataset with 0.99 precision and perfect recall. Nasr et al.~\cite{nasr2019comprehensive} suggest an adversary can actively craft his updated model to extract more membership information about other clients. For training members of the FL model, the existing inference attacks fail to explore which client owns them. The source inference attacks proposed in this paper fill this gap.

\subsection{Privacy Defenses in FL}
To enhance privacy protection, differential privacy and other privacy protection mechanisms, {\em e.g.,} secure aggregation, have been recently applied to federated learning~\cite{lyu2020privacy,bhowmick2018protection,geyer2017differentially,mcmahan2018learning,bonawitz2017practical,sun2020federated}. 
Previous works mostly focus on either the centralized differential privacy mechanism that requires a central trusted party~\cite{geyer2017differentially,mcmahan2018learning}, or local differential privacy, in which each user perturbs its updates randomly before sending it to an untrusted aggregator~\cite{truex2019hybrid,sun2020ldp}.
These privacy-preserving approaches have been evaluated effectively for inference and other attacks~\cite{geyer2017differentially,mcmahan2017learning,bonawitz2017practical,li2019fedmd,sun2020ldp} in FL.
However, no protection approaches have been explored for SIAs. As discussed in the last section, applying differential privacy in FL is not an effective solution, since it suffers from the trade-off between model utility and defense performance of SIAs, providing future research opportunities. 

\section{Conclusion}\label{7}
In this paper, we propose a new inference attack named source inference attack in the context of FL, which enables a malicious server to infer the source of a training example between clients. We derive an optimal attack strategy formally that the malicious server is able to gain non-trivial source information of the training members by evaluating the local model's loss. We evaluate SIAs in FL with many real datasets and different settings. The extensive experimental results demonstrate the effectiveness of SIAs in practice.

\bibliographystyle{abbrvnat}
\bibliography{reference}

\end{document}